\documentclass[aps,prb,showpacs,showkeys, preprint]{revtex4-1}\usepackage{amsmath}
\usepackage{amsfonts}
\usepackage[dvips]{graphicx}
\usepackage{amssymb}

\begin{document}
\title{A toy model for Macroscopic Quantum Coherence}
\author{R. Mu\~{n}oz-Vega}\email{rodrigo.munoz@uacm.edu.mx}
\affiliation{Universidad Aut\'{o}noma de la Ciudad de M\'{e}xico,
Centro Hist\'{o}rico, Fray Servando Teresa de Mier 92,
Col. Centro, Del. Cuauht\'{e}moc, M\'{e}xico D.F, C.P. 06080}
\author{Jos\'{e} Job Flores-Godoy}\email{ e-mail:job.flores@ibero.mx}
\author{ G. Fern\'{a}ndez-Anaya}\email{guillermo.fernandez@ibero.com}
\affiliation{Departamento de F\'isica y Matem\'aticas, Universidad Iberoamericana, Prol. Paseo de la Reforma 880, Col. Lomas de Santa Fe, Del. A. Obreg\'on, M\'exico, D. F. C.P. 01219}
\author{Encarnaci\'{o}n Salinas-Hern\'{a}ndez}
\email{ esalinas@ipn.mx}
\affiliation{ESCOM-IPN,
Av. Juan de Dios B\'{a}tiz s/n, Unidad Profesional Adolfo L\'{o}pez Mateos
 Col. Lindavista, Del. G. A. Madero, M\'{e}xico, D. F, C.P. 07738}
\begin{abstract}
The present article deals with Macroscopic Quantum Coherence resorting only to basic quantum mechanics. A square double well is used to illustrate the Leggett-Caldeira oscillations. The effect of thermal-radiation on two-level systems is discussed to some extent. The concept of decoherence is introduced at an elementary level. Handles are deduced for the energy, temperature and time scales involved in Macroscopic Quantum Coherence.
\end{abstract}
\date{\today}

\maketitle
\section{Introduction}
Triggered by a seminal article \cite{Leggett} written by A J Leggett in 1980, research into Macroscopic Quantum Coherence (MQC) has yielded impressive experimental,\cite{Nakamuraetal, Makhlinetal, Friedmanetal, WS} theoretical\cite{CaldeiraLeggett, Leggett1, RevDiss,  Tesche} and even technological achievements\cite{Carellietal, Manucharyanetal01, Manucharyanetal02}. The ideas developed in the last thirty so years by Leggett and his collaborators have not only changed the way we understand the relation between quantum and classical behaviours, but are also crucial in the future development of quantum computing. The present article aims at explaining the basic  phenomenology of QMC resorting only to basic quantum mechanics. Thus, we believe this article can be of interest for any student who has attended at least a one-year course in quantum physics, and for faculty members committed to introducing students into contemporary research.

In order to explain briefly what MQC is, let us consider a particle in a symmetric double well potential (SDWP). Figure 1 depicts an example of such a potential. In freshmen courses we have been told what to expect when the particle is in a high-lying energy level in a nice, analytical, potential such as this: for states for which the change in potential energy within a de Broglie wavelength is much smaller than the mean kinetic energy, the specifically quantum features of the behavior result negligible and the classical description becomes adequate.\cite{Bohm}  In that sense, classical behaviour can be considered as a limiting case of quantum mechanics \cite{LandauLif}. Suppose, nonetheless, the central barrier in the SDWP of Fig. 1 to be of macroscopic width. Then, the predictions of quantum mechanics and classical mechanics certainly clash for this system. A classical viewpoint would demand two distinct localized states of stable equilibrium, situated at $-x_{0}$ and $x_{0}$, while quantum mechanics predicts an even probability distribution for the (non-degenerate\cite{LandauA}) ground level state (which is, of course, the more stable stationary state.) In fact, the ground statefunction for such a potential would have to look something like Figure 2. 

Moreover, the (odd) eigenfunction of the first excited level (Figure 3), and indeed each one of the stationary solutions of a SDWP,  necessarily has an even probability distribution. 
%%%%%%%%%%%%%%%%%%%%%%%%%%%%%%%%%%%%%%%%%%%%%%%%%%%%%%%%%%%%%%%%%%%%%%%%%
%
%                                                                Figure figureA starts here
%
%%%%%%%%%%%%%%%%%%%%%%%%%%%%%%%%%%%%%%%%%%%%%%%%%%%%%%%%%%%%%%%%%%%%%%%%%
\begin{figure}[t]
%\label{fig:figureA}
\begin{center}
\includegraphics[angle=0, width=0.5\textwidth]{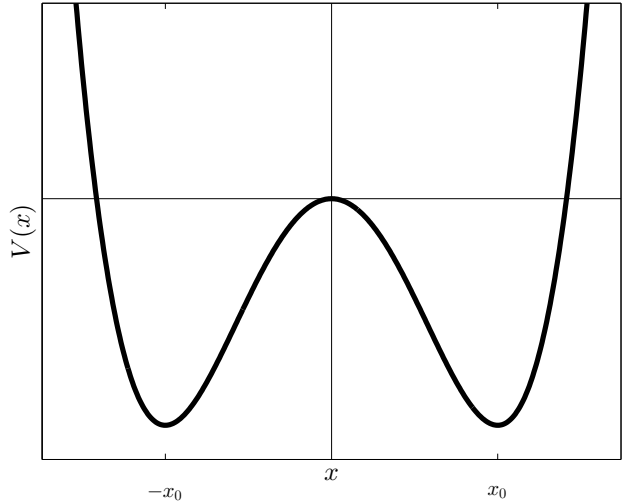}
\end{center}\caption{An example of a SDWP potential, with characteristic double minima and central peak.}
\label{fig:figure1}
\end{figure}
%%%%%%%%%%%%%%%%%%%%%%%%%%%%%%%%%%%%%%%%%%%%%%%%%%%%%%%%%%%%%%%%%%%%%%%%%
%
%                                                                Figure figureA ends here
%
%%%%%%%%%%%%%%%%%%%%%%%%%%%%%%%%%%%%%%%%%%%%%%%%%%%%%%%%%%%%%%%%%%%%%%%%%

What Leggett predicted more than thirty wears ago, and what actually happens in experiments carried out in SDWPs of micrometric and nanometric  typical lengths, is the appearance of a two-fold degenerate ground level $E^{\prime}$, with the system oscillating in an harmonic fashion between two eigenstates, $\vert L\rangle$ (Figure 4) and $\vert R \rangle$ (Figure 5) localized, respectively, at the left and right of the central barrier. At ground level, the position expectancy value oscillates in the accordance with:
\begin{equation}\label{I.1}
\langle x\rangle_{0}(t)=\langle x\rangle_{0}(0)\cos \omega t\textrm{.}
\end{equation}
This phenomenon, the so called Leggett-Caldeira oscillations, is closely related with the Rabi oscillations of atomic physics. It is explained as the result of the purported ground level $E^{\prime}$ resolving into a true ground level
\begin{equation}\label{I.2}
E_{+}=E^{\prime}-\hbar \omega/2\textrm{,}
\end{equation}
endowed with an even non-localized eigensolution $\vert +\rangle$, and a first excited level
\begin{equation}\label{I.3}
E_{-}=E^{\prime}+\hbar \omega/2\textrm{,}
\end{equation}
endowed with an odd non-localized eigensolution $\vert  -\rangle$. When a quantum system tunnels periodically trough the barrier of a SDWP with a central barrier of macroscopic length,  we have Macroscopic Quantum Coherence.

The  states $\vert R\rangle$ and $\vert L\rangle$  have, each one on its own, a definite value of a macroscopic property  (namely, the property of being localized at the left or the right of the barrier). At the same time, $\vert R \rangle$ and $\vert L\rangle$ are linear combinations of the states $\vert +\rangle$ and $\vert -\rangle$,  which cannot be said to be localized. In order to understand Leggett's original motivation, notice the analogy between macroscopic SDWPs and Schroedinger's cat: the celebrated pet can be in any of two different \emph{macroscopically distinguishable} states (let us say, $\Psi_{1}$ for a live cat and $\Psi_{0}$ for a dead one) just as a particle in a SDWP. If any of these macroscopic systems obeys the laws of quantum mechanics, then it could be prepared in linear combinations that lack a sharp, well defined, value of the macroscopic property. Examples of these linear combinations are the $\vert\pm\rangle$ states of the SDWPs, and the ``neither dead nor alive" states
\begin{equation}
\Psi_{\pm}=\frac{1}{\sqrt{2}}\Big(\Psi_{0}\pm\Psi_{1}\Big)
\end{equation}
of the cat.
%%%%%%%%%%%%%%%%%%%%%%%%%%%%%%%%%%%%%%%%%%%%%%%%%%%%%%%%%%%%%%%%%%%%%%%%%
%
%                                                                Figure B starts here
%
%%%%%%%%%%%%%%%%%%%%%%%%%%%%%%%%%%%%%%%%%%%%%%%%%%%%%%%%%%%%%%%%%%%%%%%%%
\begin{figure}[t]
\label{fig:B}
\begin{center}
\includegraphics[angle=0, width=0.5\textwidth]{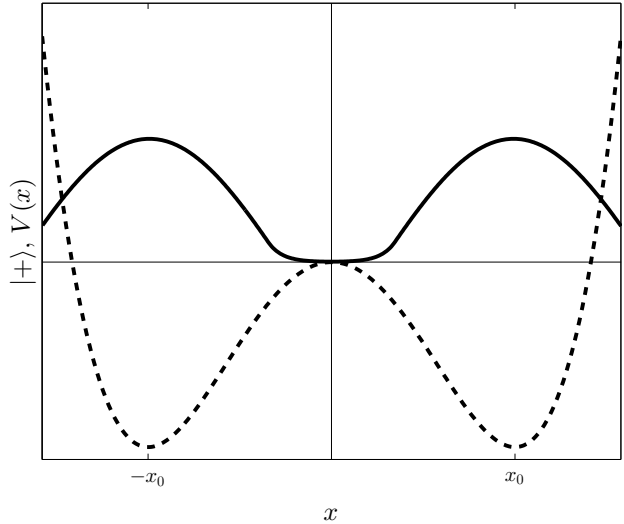}
\end{center}\caption{A rendering of what a ground-level eigenfunction (solid curve) should look like for a SDWP. The potential is shown as a dashed curve.}
\label{fig:figure1}
\end{figure}
%%%%%%%%%%%%%%%%%%%%%%%%%%%%%%%%%%%%%%%%%%%%%%%%%%%%%%%%%%%%%%%%%%%%%%%%%
%
%                                                                Figure B ends here
%
%%%%%%%%%%%%%%%%%%%%%%%%%%%%%%%%%%%%%%%%%%%%%%%%%%%%%%%%%%%%%%%%%%%%%%%%
Thus, a more general definition of MQC is simply: the quantum superposition of distinct macroscopic states. Long time before the year of 1980, macroscopic quantum phenomena had been discovered: superconductivity in 1911, and superfluidity in 1937. Yet it remained for Leggett to identify the conditions necessary for a quantum system to present macroscopically distinguishable states.\cite{Leggett}

Some twenty years elapsed between Leggett's proposal and a credible experimental confirmation\cite{Friedmanetal, WS} of  MQC. One of the main reasons for this delay lies in the fact that the phase coherence of the $\vert\pm\rangle$ states is rapidly lost due to the interaction of the system with its surroundings, so the system collapses into one of the localized states before one period of the Leggett-Caldeira oscillation is completed.\cite{Leggett, RevDiss}

MQC is relevant not only from the purely theoretical point of view. A physical qubit  is a two-level system considered as a piece of hardware. And, as we shall see in the following pages,  at least some SDWPs can behave as effective two-level systems at sufficiently low temperatures. Quantum computing (an area with impressive software development, but little hardware to show) requires qubits to interact with one another without loss of coherence, for fairly long times, even at fairly high temperatures. Thus, the study of two-level dissipative systems, to which Leggett and collaborators made far reaching contributions when delving in the foundations of quantum physics, has revealed itself crucial for people in the vanguard of technological development.\cite{Makhlinetal, WS}
%%%%%%%%%%%%%%%%%%%%%%%%%%%%%%%%%%%%%%%%%%%%%%%%%%%%%%%%%%%%%%%%%%%%%%%%%
%
%                                                                Figure C starts here
%
%%%%%%%%%%%%%%%%%%%%%%%%%%%%%%%%%%%%%%%%%%%%%%%%%%%%%%%%%%%%%%%%%%%%%%%%%
\begin{figure}[t]
\label{fig:C}
\begin{center}
\includegraphics[angle=0, width=0.5\textwidth]{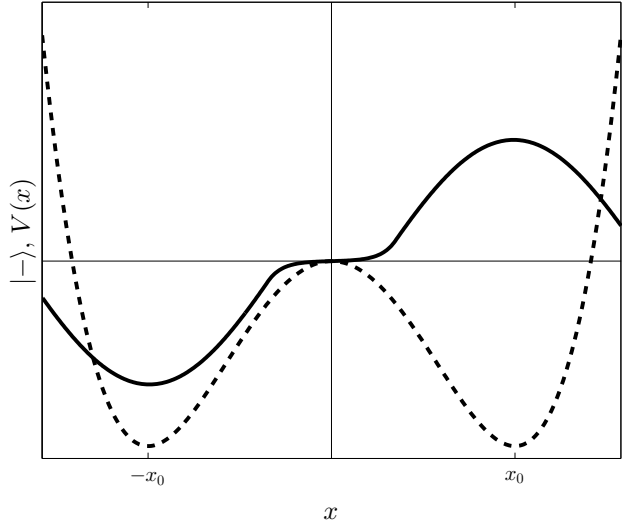}
\end{center}\caption{Th first excited level eigenfunction (solid) of a SDWP (dashed).}
\end{figure}
%%%%%%%%%%%%%%%%%%%%%%%%%%%%%%%%%%%%%%%%%%%%%%%%%%%%%%%%%%%%%%%%%%%%%%%%%
%
%                                                                Figure C ends here
%
%%%%%%%%%%%%%%%%%%%%%%%%%%%%%%%%%%%%%%%%%%%%%%%%%%%%%%%%%%%%%%%%%%%%%%%%%

The rest of this article is structured as follows: in section~\ref{sec:2} we discuss the spectra of a family of symmetric double square well potentials, and the conditions under which a member of this family can be considered as an effective two-level system. Next, the properties of two-states systems arising from SDWP´s are discussed in section ~\ref{sec:3}. We then go on to examine in ~\ref{sec:4} how thermal radiation, by throwing the system  into higher energy levels, renders the two-level model inapplicable. In section ~\ref{sec:5} decoherence is introduced in elementary terms, and its relation with dissipation is briefly discussed. Handles for the time, energy and temperature scales involved in MQC are derived from our toy model in section ~\ref{sec:6}. Finally, conclusions are laid down in section ~\ref{sec:7}. 
%%%%%%%%%%%%%%%%%%%%%%%%%%%%%%%%%%%%%%%%%%%%%%%%%%%%%%%%%%%%%%%%%%%%%%%%%
%
%                                                                            Figure D starts
%
%%%%%%%%%%%%%%%%%%%%%%%%%%%%%%%%%%%%%%%%%%%%%%%%%%%%%%%%%%%%%%%%%%%%%%%%%
\begin{figure}[t]
\label{fig:D}
\begin{center}
\includegraphics[angle=0, width=0.5\textwidth]{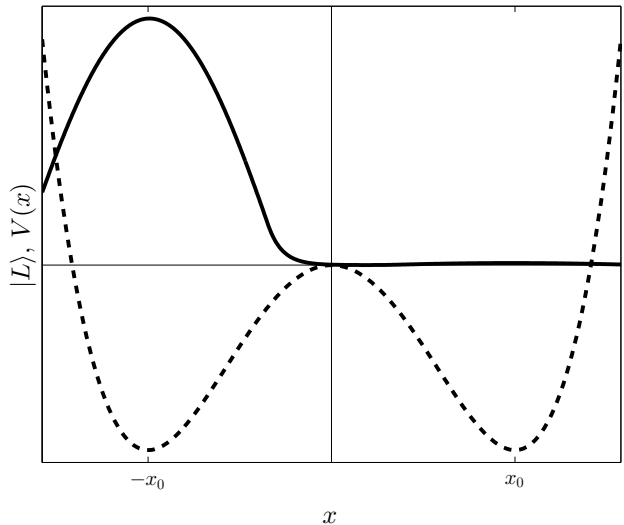}
\end{center}\caption{The $\langle x\vert L\rangle$ state (shown solid) localized at the left of the SDWP (shown dashed) barrier.}
\end{figure}
%%%%%%%%%%%%%%%%%%%%%%%%%%%%%%%%%%%%%%%%%%%%%%%%%%%%%%%%%%%%%%%%%%%%%%%%%
%
%                                                                         Figure D ends
%
%%%%%%%%%%%%%%%%%%%%%%%%%%%%%%%%%%%%%%%%%%%%%%%%%%%%%%%%%%%%%%%%%%%%%%%%%
%%%%%%%%%%%%%%%%%%%%%%%%%%%%%%%%%%%%%%%%%%%%%%%%%%%%%%%%%%%%%%%%%%%%%%%%%
%
%                                                                            Figure E starts
%
%%%%%%%%%%%%%%%%%%%%%%%%%%%%%%%%%%%%%%%%%%%%%%%%%%%%%%%%%%%%%%%%%%%%%%%%%
\begin{figure}[t]
\label{fig:E}
\begin{center}
\includegraphics[angle=0, width=0.5\textwidth]{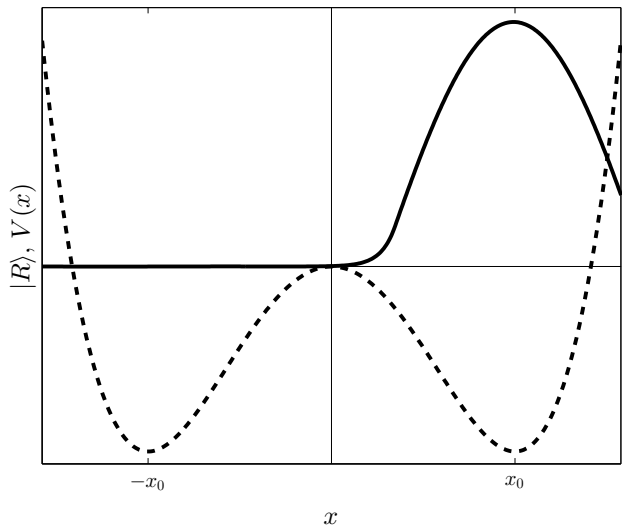}
\end{center}\caption{The $\langle x\vert R\rangle$ state (shown solid) localized at the right of the SDWP (shown dashed) barrier.}
\end{figure}
%%%%%%%%%%%%%%%%%%%%%%%%%%%%%%%%%%%%%%%%%%%%%%%%%%%%%%%%%%%%%%%%%%%%%%%%%
%
%                                                                         Figure E ends
%
%%%%%%%%%%%%%%%%%%%%%%%%%%%%%%%%%%%%%%%%%%%%%%%%%%%%%%%%%%%%%%%%%%%%%%%%%%
%
%
%%%%%%%%%%%%%%%%%%%%%%%%%%%%%%%%%%%%%%%%%%%%%%%%%%%%%%%%%%%%%%%%%%%%%%%%%%%
%
%
%                                                                                                   Section 2 starts here
%
%
%%%%%%%%%%%%%%%%%%%%%%%%%%%%%%%%%%%%%%%%%%%%%%%%%%%%%%%%%%%%%%%%%%%%%%%%%%%
\section{Symmetric Double Square Wells}\label{sec:2}
Leggett resorted to quasi-classical considerations when stating his original proposal.\cite{Leggett} Also, the WKB approximation has been applied to double well potentials  by Landau and Lifshitz,\cite{LLandau} and more recently, in this Journal,  by others.\cite{JelicMarsiglio} Here will take a different point of view, avoiding all together quasi-classical approximations, by considering a particular family of double infinite square well potentials as approximations to actual, analytic SDWPs. Our procedure will later allow us to get some reference values on the energies, temperatures and times involved in MQC. The following family of piece-wise-constant potentials will be considered:
\begin{equation}\label{ucases}
U_{b}(x)=\left \{ \begin{array}{c l}
\infty &\textrm{if } x \leq -a-b\textrm{,}\\
0 &\textrm{if } -b > x > -a-b\textrm{,}\\
k &\textrm{if } b\geq x \geq -b\textrm{,}\\
0 &\textrm{if } b+a> x > b\textrm{,}\\
\infty &\textrm{if } x \geq b+a\textrm{.}\\
\end{array}\right .
\end{equation}
Potentials of this kind have previously been studied in a different context, and it has been shown\cite{Munoz}  that, if  all other parameters held fixed,  levels $E_{2n+1}$ and $E_{2n}$ coalesce as $k\rightarrow \infty$. Here we shall consider the barrier height  $k>0$ as a fixed number, although ``big" in a sense that will be readily clarified. This, in order to keep the gap between the ground and first excited levels sufficiently small. We shall also take the width of each one of the lateral valleys,  $a>0$, as a fixed value unless otherwise stated, leaving free the only other parameter, that is the barrier half-width $b>0$.

One of the two main objectives of this Section is to obtain a global lower bound on the energy gap between the first and second excited levels in the $U_{b}$ potentials. Just as important to our ends, we will learn on this Section that there is a ``running" upper bound (dependent on the value of $b$) on the gap between the ground and first excited levels. The consequences of this to facts, which are vital to the rest of the article, are explored in Sections III, IV and VI. 

 To be sure, non of the $U_{b}$ is continuous, yet they share the most prominent features of a SDWP, namely, they are even potentials with completely bounded, non-degenerate, spectra, as can be shown from boundary conditions. If instead of two minima, the $U_{b}$ have two non-overlapping regions of minima (\emph{viz.} $(-a-b,-b)$ and $(b,a+b)$), this distinction will prove to be quite unimportant. 
%%%%%%%%%%%%%%%%%%%%%%%%%%%%%%%%%%%%%%%%%%%%%%%%%%%%%%%%%%%%%%%%%%%%%%%%%%
%
%                                                                                Figure F starts
%
%
%%%%%%%%%%%%%%%%%%%%%%%%%%%%%%%%%%%%%%%%%%%%%%%%%%%%%%%%%%%%%%%%%%%%%%%%%
\begin{figure}[t]
\label{msbs1}
\begin{center}
\includegraphics[angle=0, width=0.5\textwidth]{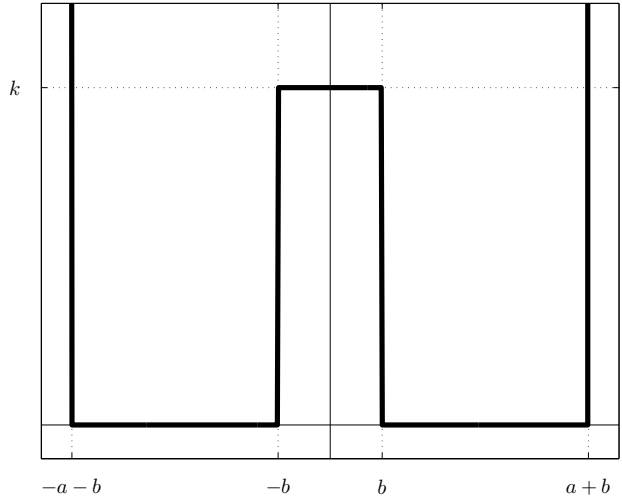}
\end{center}
\caption{A typical member of the $U_b(x)$ family of potentials}
\label{fig:figure5r}
\end{figure}
%%%%%%%%%%%%%%%%%%%%%%%%%%%%%%%%%%%%%%%%%%%%%%%%%%%%%%%%%%%%%%%%%%%%%%%%%%
%
%                                                                                 Figure F ends
%
%
%%%%%%%%%%%%%%%%%%%%%%%%%%%%%%%%%%%%%%%%%%%%%%%%%%%%%%%%%%%%%%%%%%%%%%%%%

Also from boundary conditions (or from more abstract, symmetry considerations)  it is readily seen that the levels in the spectrum of any of the  $U_{b}$ are classified according with parity, just like it happens for a continuous even potential:
\begin{equation}\label{II.1}
\psi_{2n,b}(-x)=\psi_{2n,b}(x)\textrm{,}\quad\quad n=0,1,\ldots\ \textrm{,}\quad b\in (0, \infty )
\end{equation}
and
\begin{equation}\label{II.2}
\psi_{2n+1,b}(-x)=-\psi_{2n+1,b}(x)\textrm{,}\quad\quad n=0,1,\ldots\ \textrm{,}\quad b\in (0, \infty )\textrm{.}
\end{equation}

Let us focus on the discretization conditions below the level of the central barrier ($E<k$). From the boundary conditions we get, for even states:
\begin{equation}\label{coneven}
-\sqrt{E_{2n}}\cot a\frac{\sqrt{2mE_{2n}}}{\hbar}=\sqrt{k-E_{2n}}\tanh b\frac{\sqrt{2m(k-E_{2n})}}{\hbar}\quad\textrm{,}
\end{equation}
while odd levels below the barrier level have to comply with
\begin{equation}\label{conodd}
-\sqrt{E_{2n+1}}\cot a\frac{\sqrt{2mE_{2n+1}}}{\hbar}=\sqrt{k-E_{2n+1}}\coth b\frac{\sqrt{2m(k-E_{2n+1})}}{\hbar}\quad\textrm{.}
\end{equation}
Notice how the first of these two conditions can be written in the form:
\begin{equation}
g(E_{2n})=h_{b}(E_{2n}),
\end{equation}
and the second can be rendered as:
\begin{equation}
g(E_{2n+1})=j_{b}(E_{2n+1}),
\end{equation}
with the meaning of $g, h_{b}$ and $j_{b}$ being obvious from the context.

Both of these two last equations are depicted in Figure 7, from which it can be seen that there exists an upper  bound $B$, given by
\begin{equation}\label{inequalityA}
B=\frac{\pi^{2}\hbar ^{2}}{2m a^{2}},
\end{equation}
such that the ground and first excited states have to comply with
\begin{equation}\label{funnyeq}
\frac{B}{4}<E_{0}<E_{1}<B ,
\end{equation}
no matter the value of $b.$ Obviously, there can be no levels below the barrier unless $k>B/4$. We shall only consider potentials for which the condition:
\begin{equation}\label{bigenough}
k>>B
\end{equation}
is met, so that we will always  have at  least two levels below the barrier. Indeed, the number of levels below the barrier increases with increasing quotient $k/B$ and, more importantly, as $B$ is independent of $k$, condition (\ref{bigenough}) warrants that the gap between the first two level is always small. 
It is not difficult to generalize (\ref{funnyeq}) starting from  (\ref{coneven}) and (\ref{conodd}) and definition (\ref{inequalityA}). The result is that:
\begin{equation}\label{III.1.1}
(n+\frac{1}{2})^{2}B<E_{2n,k}<E_{2n+1,k}<(n+1)^{2}B\textrm{,}\quad n=0,1,2,\ldots N,
\end{equation}
if the level $2N+1$ is still below the barrier.

From inequality (\ref{III.1.1}) it follows that
\begin{equation}\label{III.1.2}
E_{2n+2}-E_{2n+1}>(n+5/4)B\textrm{,}\quad n=0,1,2,\ldots, N
\end{equation}
if level $2N+1$ is below the barrier. We then have that the gap between the ground and first excited levels will always be less than the gap between the first and second excited levels:
\begin{equation}\label{inek01}
E_{2}-E_{1}>\frac{5}{4}B>\frac{3}{4}B> E_{1}-E_{0}.
\end{equation}
But we can do much more better than that. Indeed, in Appendix A it is formally proven that for any given number $\delta>0$ there exist a value $b(\delta)>0$ such  the gap between the ground and the first excited level of a $U_{b}$ potential will be less than $\delta$, that is
\begin{equation}
E_{1}-E_{0}<\delta ,
\end{equation}
if $b\geq b(\delta)$. In other words, if we choose the barrier length big enough, then we can make $E_{0}$ and $E_{1}$ as proximate as we want, while there is a lower bound for the gap between $E_{2}$ and $E_{1}$ which is independent of the value of this length. This will allow us to find examples of $U_{b}$ that will work as effective two-state systems for the lowest-lying energy levels, as illustrated in Figure 8.
%%%%%%%%%%%%%%%%%%%%%%%%%%%%%%%%%%%%%%%%%%%%%%%%%%%%%%%%%%%%%%%%%%%%%%%%%%
%
%                                                                                Figure G starts
%
%
%%%%%%%%%%%%%%%%%%%%%%%%%%%%%%%%%%%%%%%%%%%%%%%%%%%%%%%%%%%%%%%%%%%%%%%%%
\begin{figure}[t]
%\label{msbs1}
\begin{center}
\includegraphics[angle=0, width=0.5\textwidth]{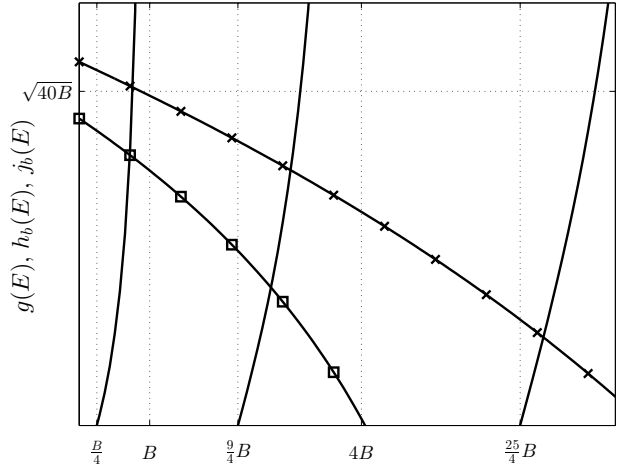}
\end{center}\caption{Graphical solutions of transcendental equations (\ref{coneven}) and (\ref{conodd}). Depicted, functions $g(E)$ (solid), $h_{b}(E)$ (squares) and $j_{b}(E)$ (crosses). In all cases $k=40B$ and $b=0.8a$. In this example only the ground level and the three first excited levels are below the barrier height $k$.}
\end{figure}
%%%%%%%%%%%%%%%%%%%%%%%%%%%%%%%%%%%%%%%%%%%%%%%%%%%%%%%%%%%%%%%%%%%%%%%%%%
%
%                                                                                Figure G ends
%
%
%%%%%%%%%%%%%%%%%%%%%%%%%%%%%%%%%%%%%%%%%%%%%%%%%%%%%%%%%%%%%%%%%%%%%%%%%

%%%%%%%%%%%%%%%%%%%%%%%%%%%%%%%%%%%%%%%%%%%%%%%%%%%%%%%%%%%%%%%%%%%%%%%%%%
%
%                                                                                Figure H starts
%
%
%%%%%%%%%%%%%%%%%%%%%%%%%%%%%%%%%%%%%%%%%%%%%%%%%%%%%%%%%%%%%%%%%%%%%%%%%
\begin{figure}[t]
%\label{msbs1}
\begin{center}
\includegraphics[angle=0, width=0.5\textwidth]{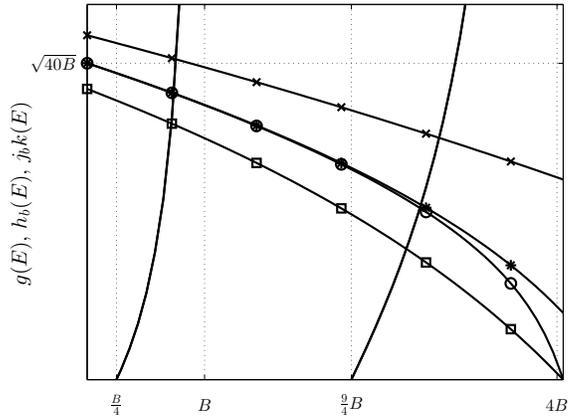}
\end{center}\caption{Doubling the barrier width produces a dramatic decrease in the gap between the ground and first excited levels. Shown: function $h_{b}$ for $b=0.8a$ (squares) and $b=0.4a$(circles), and function $j_{b}$ for $b=0.8a$ (crosses) and $b=0.4a$ (asterisks). In all cases $k=40B.$}
\end{figure}
%%%%%%%%%%%%%%%%%%%%%%%%%%%%%%%%%%%%%%%%%%%%%%%%%%%%%%%%%%%%%%%%%%%%%%%%%%
%
%                                                                                Figure H ends
%
%
%%%%%%%%%%%%%%%%%%%%%%%%%%%%%%%%%%%%%%%%%%%%%%%%%%%%%%%%%%%%%%%%%%%%%%%%%

Finally, there is one more inequality that can be derived from (\ref{III.1.1}) and that will prove useful in section IV. This inequality is:
\begin{equation}\label{2NF}
E_{2}-E_{1}<\frac{15}{4}B\ .
\end{equation}
Let us stress that relations (\ref{funnyeq}), (\ref{inek01}) and (\ref{2NF}) are verified for each $U_{b}$ regardless of the value of $b.$ 
%%%%%%%%%%%%%%%%%%%%%%%%%%%%%%%%%%%%%%%%%%%%%%%%%%%%%%%%%%%%%%%%%%%%%%%%%%
%
%                                                                                Figure I begins
%
%
%%%%%%%%%%%%%%%%%%%%%%%%%%%%%%%%%%%%%%%%%%%%%%%%%%%%%%%%%%%%%%%%%%%%%%%%%
\begin{figure}[t]
\begin{center}
\includegraphics[angle=0, width=0.5\textwidth]{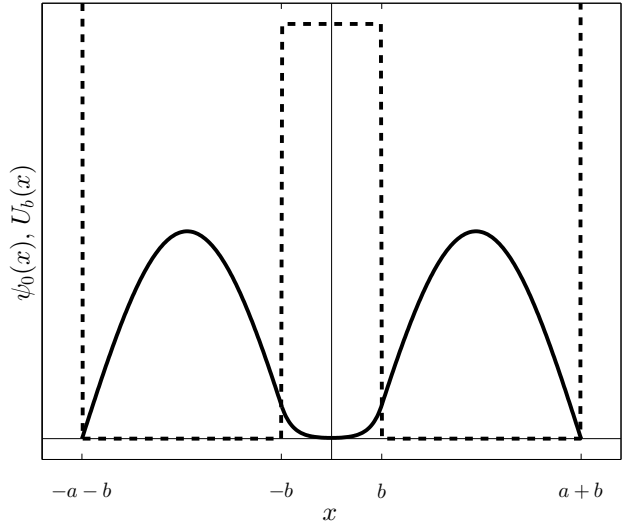}
\end{center}\caption{Solid curve: the $\psi_{+}(x)$ ground-level eigenfunction (even), obtained through computer assisted numerical analysis. Dashed: the corresponding $U_{b}$ potential. In this example $k=14B$ and $b=0.$   }
\label{fig:figure1}
\end{figure}
%%%%%%%%%%%%%%%%%%%%%%%%%%%%%%%%%%%%%%%%%%%%%%%%%%%%%%%%%%%%%%%%%%%%%%%%%%
%
%                                                                                Figure I ends
%
%
%%%%%%%%%%%%%%%%%%%%%%%%%%%%%%%%%%%%%%%%%%%%%%%%%%%%%%%%%%%%%%%%%%%%%%%%%

%%%%%%%%%%%%%%%%%%%%%%%%%%%%%%%%%%%%%%%%%%%%%%%%%%%%%%%%%%%%%%%%%%%%%%%%%
%
%
%                                                                              Section 3 begins
%
%
%%%%%%%%%%%%%%%%%%%%%%%%%%%%%%%%%%%%%%%%%%%%%%%%%%%%%%%%%%%%%%%%%%%%%%%%%%
\section{Two-level systems with reflection symmetry}\label{sec:3}
In the preceding section we have proven that there are $U_{b}$ potentials for which the gap between the ground and first excited energy  levels is much more narrow than the one between the first and second excited levels. Consequently, for low energy expectancy values, a particle in one of such potentials acts as an effective two-level systems.\cite{Feynman, Cohen}

In the rest of this section we shall consider a fixed $U_{b}$ that behaves as a two-level system, and drop the $b$.

Consider now the non-stationary solutions $\psi_{L}$ and $\psi_{R}$ that one obtains from the linear combinations
\begin{equation}\label{II.3}
\psi_{L} (x,t)=\frac{1}{\sqrt{2}}\Big [\exp \left(-\imath\frac{E_{0}t}{\hbar}\right) \psi_{0}(x) + \exp\left(-\imath\frac{E_{1}t}{\hbar}\right)\psi_{1}(x)\Big ]
\end{equation}
%%%%%%%%%%%%%%%%%%%%%%%%%%%%%%%%%%%%%%%%%%%%%%%%%%%%%%%%%%%%%%%%%%%%%%%%%
%
%                                                                 Figure I begins
%
%%%%%%%%%%%%%%%%%%%%%%%%%%%%%%%%%%%%%%%%%%%%%%%%%%%%%%%%%%%%%%%%%%%%%%%%%
\begin{figure}[t]
%\label{msbs1}
\begin{center}
\includegraphics[angle=0, width=0.5\textwidth]{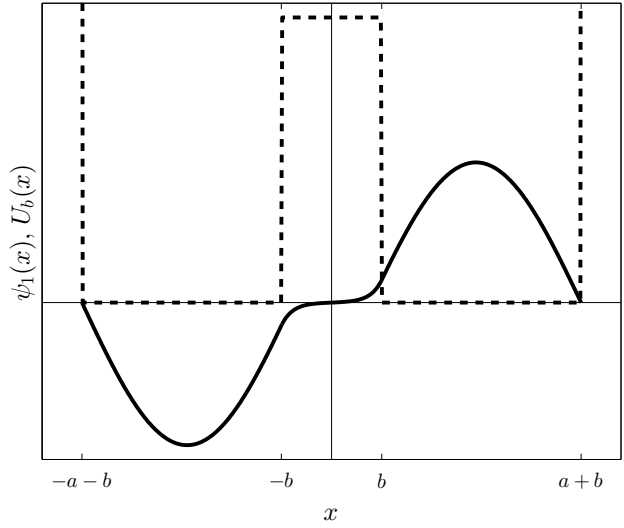}
\end{center}\caption{ Solid: the (odd) eigenfunction of the first excited level. Dashed: the $U_{b}$ potential to which this solution corresponds. In this example $k=14B$ and $b=0.2a$.}
\end{figure}
%%%%%%%%%%%%%%%%%%%%%%%%%%%%%%%%%%%%%%%%%%%%%%%%%%%%%%%%%%%%%%%%%%%%%%%%%%
%
%                                                                    Figure I ends
%
%%%%%%%%%%%%%%%%%%%%%%%%%%%%%%%%%%%%%%%%%%%%%%%%%%%%%%%%%%%%%%%%%%%%%%%%%
and
\begin{equation}\label{II.4}
\psi_{R} (x,t)=\frac{1}{\sqrt{2}}\Big [ \exp\left(-\imath\frac{E_{0}t}{\hbar}\right) \psi_{0}(x) - \exp\left(-\imath\frac{E_{1}t}{\hbar}\right) \psi_{1}(x)\Big]\textrm{.}
\end{equation}
These states have no definite parity, but instead one is the specular image of the other:
\begin{equation}\label{II.A.1}
\psi_{L}(-x,t)=\psi_{R}(x,t)\textrm{,}
\end{equation}
as can be seen from equations (\ref{II.1}), (\ref{II.2}), (\ref{II.3}) and (\ref{II.4}).

The position expectancy value for this states is calculated from  (\ref{II.1}) in a straightforward manner:
\begin{equation}\label{II.5}
\langle x\rangle_{L}(t) =\ -\ \langle x\rangle_{R}(t) =\ \langle\psi_{0}\vert x\vert \psi_{1}\rangle\cos\frac{E_{1}-E_{0}}{\hbar}t,
\end{equation}
as is the energy expectancy value:
\begin{equation}\label{II.6}
\langle H\rangle_{L}=\langle H \rangle_{R}=\frac{E_{0}+E_{1}}{2}\textrm{.}
\end{equation}
Comparing (\ref{II.5}) with (\ref{I.1}) and (\ref{II.6}) with (\ref{I.2}) one may be tempted to make the identifications
\begin{equation}\label{II.7}
E^{\prime}=\langle H\rangle_{L}\quad\textrm{ and }\quad \omega=\frac{E_{1}-E_{2}}{\hbar}\textrm{,}
\end{equation}
from which (\ref{I.3}) would follow, so that the states of (\ref{II.3}) and (\ref{II.4}) could be interpreted as the localized states observed in the experiments, and $\psi_{0}$ and $\psi_{1}$ would correspond to the true ground level $E_{+}$ and the first excited state $E_{-}$. That is, it would be cogent that
\begin{equation}
\langle x\vert L\rangle=\psi_{L}(x)\textrm{,}\quad\langle x\vert R\rangle=\psi_{R}(x)\textrm{,}\quad
\langle x\vert +\rangle=\psi_{0}(x)\textrm{,}\quad\langle x\vert -\rangle=\psi_{1}(x)\textrm{.}
\end{equation}
%%%%%%%%%%%%%%%%%%%%%%%%%%%%%%%%%%%%%%%%%%%%%%%%%%%%%%%%%%%%%%%%%%%%%%%%%%
%
%                                                                    Figure J begins
%
%%%%%%%%%%%%%%%%%%%%%%%%%%%%%%%%%%%%%%%%%%%%%%%%%%%%%%%%%%%%%%%%%%%%%%%%%
\begin{figure}[t]
\begin{center}
\includegraphics[angle=0, width=0.5\textwidth]{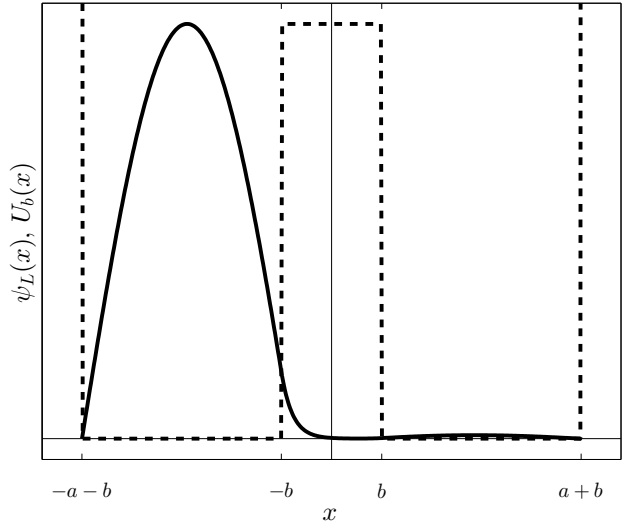}
\end{center}\caption{The localized $\psi_{L}(x,t=0)$ eigenfunction (solid) of the $U_{b}$ potential (dashed) for $b=0.4a$ and $k=14B$.}
\end{figure}
%%%%%%%%%%%%%%%%%%%%%%%%%%%%%%%%%%%%%%%%%%%%%%%%%%%%%%%%%%%%%%%%%%%%%%%%%%
%
%                                                                    Figure J ends
%
%%%%%%%%%%%%%%%%%%%%%%%%%%%%%%%%%%%%%%%%%%%%%%%%%%%%%%%%%%%%%%%%%%%%%%%%%
In this interpretation, however, there is no room for transitions. Indeed, the complete Schroedinger equation for a $U$ potential, which reads:
\begin{equation}\label{SchU}
-\frac{\hbar^{2}}{2m}\frac{\partial^{2}\psi}{\partial x^{2}}(x,t)+U(x)\psi(x,t)=i\hbar\frac{\partial\psi}{\partial t}(x,t) \textrm{,}
\end{equation}
predicts that if the system is initially prepared in the state $\psi_{L}(x,t=0)$ at time $t=0$, then it will remain in the $\psi_{L}(x,t)$ state for $t\in[0,\infty)$ (which is a sophisticated way to say: \emph{forever}). This is just consequence of PDE's theory.

Instead of periodical transitions between two different states, our equations predict the existence of a unique ``oscillating" state, because $\psi_{R}(x,t)$ is a time-displaced replica of $\psi_{L}(x,t)$:
\begin{figure}[t]
\begin{center}
\includegraphics[angle=0, width=0.5\textwidth]{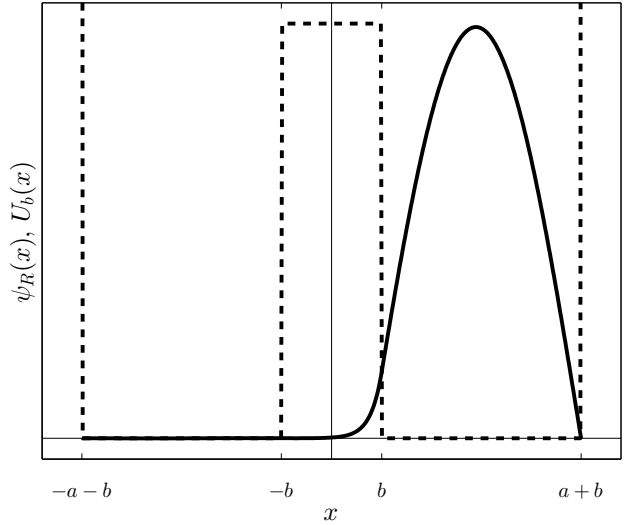}
\end{center}\caption{The localized $\psi_{R}(x,t=0)$ eigenfunction (solid) of the $U_{b}$ potential (dashed) for $b=0.4a$ and $k=14B$.}
\end{figure}
\begin{equation}\label{II.F}
\psi_{L}\Big(x,\ t+\frac{\pi}{\omega}\Big)=\imath\exp\Big(-\imath\frac{\pi\Omega}{\omega}\Big)\ \psi_{R}(x,t) \quad\textrm{ ,}
\end{equation}
where $\Omega$ stands for
\begin{equation}\label{II.G}
\Omega=(E_{1}+E_{2})/2\hbar
\end{equation}
and $\omega$ is as in (\ref{II.7}). One arrives at this result directly from (\ref{II.3}) and (\ref{II.4}) after some algebra.
\subsection{Flip-flops and Leggett-Caldeira oscillations}
Let us start from what we know happens in actual experiments (\emph{i. e.} the existence of an observable degenerate ground level) and proceed to deduce from there the perturbation needed to achieve such degeneracy. The SDWP Hamiltonian $H$ is represented by the matrix
\begin{equation}\label{IV.A.a}
\mathbb{H}=
\left(\begin{array}{cc}
E_{0}&0\\
0&E_{1}\\
\end{array}\right)
\end{equation}
in the symmetry-respecting basis formed by the eigenfunctions $\psi_{0}$ and $\psi_{1}$. Let us consider another Hamiltonian, $H^{\prime}$,  represented by the matrix
\begin{equation}\label{IV.B.4}
\tilde{\mathbb{H}}^{\prime}=\mathbb{O}\mathbb{H}^{\prime}\mathbb{O}^{-1}=
\left(\begin{array}{cc}
E^{\prime}&0\\
0&E^{\prime}\\
\end{array}\right)
\end{equation}
in the symmetry-violating basis spanned by $\psi_{L}$ and $\psi_{R}$. Here, $\mathbb{O}$ stands for the unitary operator which transforms $\psi_{0}$ into $\psi_{L}$ and $\psi_{1}$ into $\psi_{R}$, thus:
\begin{equation}\label{IV.A.5}
\mathbb{O}\left(\begin{array}{c}
1\\
0\\
\end{array}\right)=\frac{1}{\sqrt{2}}\left(\begin{array}{c}
1\\
1\\
\end{array}\right)
\quad\textrm{ and }\quad
\mathbb{O}\left(\begin{array}{c}
0\\
1\\
\end{array}\right)=
\frac{1}{\sqrt{2}}\left(\begin{array}{c}
1\\
-1\\
\end{array}\right).
\end{equation}
Notice that, as the rhs of (\ref{IV.B.4}) is proportional to the identity matrix, then:
\begin{equation}
\mathbb{H}^{\prime}=\left(\begin{array}{cc}
E^{\prime}&0\\
0&E^{\prime}\\
\end{array}\right).
\end{equation}
Thus, the perturbation is represented by:
\begin{equation}
\mathbb{W}=\mathbb{H}^{\prime}-\mathbb{H}=\left(\begin{array}{cc}
\hbar \omega/2&0\\
0&-\hbar \omega/2\\
\end{array}\right).
\end{equation}
Now, in the basis spanned by $\psi_{L}$ and $\psi_{R}$ the things look quite different. Indeed, we have that:
\begin{equation}
\tilde{\mathbb{H}}=\mathbb{OHO}^{-1}=\left(\begin{array}{cc}
E^{\prime}&-\hbar \omega/2\\
-\hbar \omega/2&E^{\prime}\\
\end{array}\right)
\end{equation}
and most importantly:
\begin{equation}
\tilde{\mathbb{W}}=\mathbb{OWO}^{-1}=\left(\begin{array}{cc}
0&\hbar \omega/2\\
\hbar \omega/2&0\\
\end{array}\right).
\end{equation}
So that the perturbation has no diagonal elements. This means that zeroth order corrections are strictly null for the perturbation, and, further more, that the off-diagonal elements are equal.

To be very clear, let us write the eigen-equations for each one of this distinct systems. For $H$ we have
\begin{equation}\begin{array}{cc}
H\psi_{0}(x,t)=i\hbar\frac{\partial\psi_{0}}{\partial t}(x,t)=E_{0}\psi_{0}(x,t),&H\psi_{1}(x,t)=i\hbar\frac{\partial\psi_{1}}{\partial t}(x,t)=E_{1}\psi_{1}(x,t),\\
\end{array}
\end{equation}
while $H^{\prime}$ responds to
\begin{equation}\label{IV.Z}
\begin{array}{cc}
H^{\prime}\psi_{L}(x,t)=i\hbar\frac{\partial\psi_{L}}{\partial t}(x,t)=E^{\prime}\psi_{L}(x,t),&H^{\prime}\psi_{R}(x,t)=i\hbar\frac{\partial\psi_{R}}{\partial t}(x,t)=E^{\prime}\psi_{R}(x,t).\\
\end{array}
\end{equation}
Now, let us consider $\tilde{\mathbb{H}}^{\prime}$ as the initial, unperturbed, Hamiltonian matrix, and
\begin{equation}
-\tilde{\mathbb{W}}=-\mathbb{OWO}^{-1}
\end{equation}
as the perturbation, so that $\tilde{\mathbb{H}}$ is  the final, perturbed, Hamiltonian matrix.  Then we can show that the $\psi_{L}(x,t)$ and $\psi_{R}(x,t)$ states transit from one another in Rabi style. Indeed, resorting to the time-dependent perturbation formalism,\cite{Landau2, Fitzpatrick1} we write, for a general state $\psi(x,t)$ of $\tilde{\mathbb{H}},$
\begin{equation}\label{IV.A.7}
\psi(x,t)=c_{L}(t)\psi_{L}(x,t)+c_{R}(t)\psi_{R}(x,t),
\end{equation}
in order to obtain the equation
\begin{equation}\label{mtdpC}
i\hbar\frac{d}{dt}\left(\begin{array}{c}
c_{0}\\
c_{1}\\
\end{array}\right)(t)=\tilde{\mathbb{W}}\left(\begin{array}{c}
c_{0}\\
c_{1}\\
\end{array}\right)(t),
\end{equation}
which is equivalent to the $2\times 2$ system of coupled linear equations:
\begin{equation}\label{IV.A.8}
i\hbar\frac{dc_{L}}{dt}=-\frac{\hbar \omega}{2}c_{R}\quad , i\hbar\frac{dc_{R}}{dt}=-\frac{\hbar \omega}{2}c_{L}.\\
\end{equation}
By uncoupling this system we get the harmonic oscillator equation
\begin{equation}
\frac{d^{2}c_{L}}{dt^{2}}=-\frac{\omega^{2}}{4}c_{L}
\end{equation}
and a similar equation for $c_{R}$, so that
\begin{equation}\label{Rabires}
c_{L}(t)=\sin\big(\omega t/2+\phi\big)\quad c_{R}(t)=\cos\big(\omega t/2+\phi\big),
\end{equation}
where $\phi$ is a constant that can be elucidate from initial conditions. The probability of finding the particle in the state $\psi_{L}$ is given, according to this last equations, by
\begin{equation}\label{probell}
P_{L}(t)=\sin^{2}\big(\omega t/2+\phi\big),
\end{equation}
and the probability of finding the the particle in the $\psi_{R}$ state is
\begin{equation}\label{probar}
P_{R}(t)=1-P_{L}(t).
\end{equation}
This is a particular instance of the Rabi oscillation, and this case is resonant due to the degeneracy of the ``initial" Hamiltonian $\tilde{H}^{\prime}$. But the ``perturbed" Hamiltonian $H$ is nothing else than the SDWP Hamiltonian of equation (\ref{SchU}).

Now, equations (\ref{probell}) and (\ref{probar}) predict the ``flip-flop" between the stationary states $\psi_{R}(x)$ and $\psi_{L}(x)$, so that, if the system is initial prepared in the state
\begin{equation}
\psi(x,t=0)=\psi_{L}(x),
\end{equation}
then we will have a 100\% certainty to find it in state $\psi_{R}(x)$ at times $t=\frac{\pi}{\omega},\frac{3\pi}{\omega},\frac{5\pi}{\omega}\ldots$ and a 100\% certainty to find it in state $\psi_{L}(x)$ at times  $t=\frac{2\pi}{\omega},\frac{4\pi}{\omega},\frac{6\pi}{\omega}\ldots$.  And this last result is consistent with equation (\ref{II.F}). Thus, we are in the presence of two different (yet not contradictory) descriptions of one and the same phenomenon: if $H$ is considered an unperturbed Hamiltonian, with complete stationary solutions $\psi_{0}(x,t)$ and $\psi_{1}(x,t)$, then we have an ``oscillating" non-stationary solution $\psi_{L}(x,t)$. If, on the other hand, $H$ is considered to be the result of a perturbation acting on the degenerate Hamiltonian $H^{\prime}$, we then get flip-flops between the complete stationary solutions of $H^{\prime}$, that is: $\psi_{L}(x,t)$ and $\psi_{R}(x,t)$.

In this manner, we obtain the periodic transitions (the zero point Leggett-Caldeira oscillations) observed in so many experiments. Notice that this transitions occur in the absence of external fields, thus without emission or absorption.
%%%%%%%%%%%%%%%%%%%%%%%%%%%%%%%%%%%%%%%%%%%%%%%%%%%%%%%%%%%%%%%%%%%%%%%%
%
%
%                                                                                                Section 4 begins
%
%
%%%%%%%%%%%%%%%%%%%%%%%%%%%%%%%%%%%%%%%%%%%%%%%%%%%%%%%%%%%%%%%%%%%%%%%%%
\section{Thermal radiation}\label{sec:4}
Due to the fact that  no quantum system can be completely isolated from its environment, in any realistic description the Schroedinger equation must be supplemented with terms that describe the interaction between the system and its surroundings. But there are very different ways to  describe this interaction and its results, depending on the time and energy scales  involved, and the complexity of the analysis. Here we shall discuss the absorption-induced transitions by which the system is thrown into high-lying energy levels, rendering the two-level model inapplicable. The main result from this discussion will be a limit on the temperature at which  Caldeira-Leggett oscillations can be observed.
\subsection{Oscillations near resonance}
Now, oscillatory behavior is to be expected not only for the the resonant, exactly degenerate, Hamiltonian matrix $\mathbb{H}^{\prime}$. Indeed, it would not be realistic to expect Leggett-Caldeira oscillations only in perfectly isolated systems.  Consider a harmonic perturbation of the SDWP matrix  Hamiltonian $\mathbb{H}$ of equation (\ref{IV.A.a}), that is, a perturbative term of the general form
\begin{equation}
\mathbb{V}=\mathbb{A}\exp(i\omega^{\prime}t)+\mathbb{A}^{\dagger}\exp(-i\omega^{\prime}t),
\end{equation}
and let us focus on the particularly simple case for which
\begin{equation}\label{simple.1}
\mathbb{A}=A\left(\begin{array}{cc}
0&1\\
0&0\\
\end{array}\right),
\end{equation}
so that the perturbative term can be written down as
\begin{equation}
\mathbb{V}=A\left(\begin{array}{cc}
0&\exp(i\omega^{\prime}t)\\
\exp(-i\omega^{\prime}t)&0\\
\end{array}\right).
\end{equation}
We then again resort to the time-dependent perturbation formalism, and write
\begin{equation}
\psi(x,t)=c_{0}(t)\psi_{0}(x,t)+c_{1}(t)\psi_{1}(x,t)
\end{equation}
in order to obtain the equation
\begin{equation}\label{mtdpC}
i\hbar\frac{d}{dt}\left(\begin{array}{c}
c_{0}\\
c_{1}\\
\end{array}\right)(t)=\mathfrak{V}(t)\left(\begin{array}{c}
c_{0}\\
c_{1}\\
\end{array}\right)(t),
\end{equation}
where $\mathfrak{V}$, defined by
\begin{equation}
\mathfrak{V}(t)=\exp\Big(i\mathbb{H} t/\hbar\Big)\mathbb{V}(t)\exp\Big(-i\mathbb{H} t/\hbar\Big)
\end{equation}
represents the perturbation in the interaction picture, and in our particularly simple case reduces to
\begin{equation}
\mathfrak{V}(t)=A\left(\begin{array}{ccc}
0&\quad&\exp \left(i(\omega^{\prime}-\omega)t \right)\\
 & \\
\exp \left(-i(\omega^{\prime}-\omega)t \right) &\quad&0\\
\end{array}\right),
\end{equation}
so that equation (\ref{mtdpC}) is equivalent to the $2 \times 2$ system of coupled ODEs
\begin{equation}
i\hbar\frac{dc_{0}}{dt}=A\exp\Big[ i\big(\omega^{\prime}-\omega\big)t\Big] c_{1}\textrm{,}\quad
i\hbar\frac{dc_{1}}{dt}=A\exp\Big[ -i\big(\omega^{\prime}-\omega\big)t\Big] c_{0}\textrm{.}
\end{equation}
It can be checked by hand that
\begin{equation}
c_{0}(t)=\exp(i\Omega^{\prime} t/2)\Bigg\{\cos (R_{0} t)-\frac{i\Omega^{\prime}}{2R_{0}}\sin (R_{0}t)\Bigg\}
\end{equation}
and
\begin{equation}
c_{1}(t)=\frac{-iR_{1}}{R_{0}}\exp(-i\Omega^{\prime} t/2)\sin (R_{0}t)
\end{equation}
provide a solution for the initial conditions $c_{0}(t=0)=1 , c_{1}(t=0)=0$. Here we have used the following shorthand
\begin{equation}\begin{array}{cccc}
R_{0}=\sqrt{(A/\hbar)^{2}+\Big(\frac{\omega^{\prime}-\omega}{2}\Big)^{2}},&\ \Omega^{\prime}=\omega^{\prime}-\omega &
\textrm{ and }&
R_{1}=A/\hbar,\
\end{array}
\end{equation}
which lead to what is known as Rabi's formula,\cite{Fitzpatrick} namely:
\begin{equation}\label{Rah.1}
P_{1}(t)=\bigg(\frac{R_{1}}{R_{0}}\bigg)^{2}\sin^{2}\big(R_{0}t\big),
\end{equation}
\begin{equation}\label{Rah.2}
P_{0}(t)=1-P_{1}(t).
\end{equation}
It is not difficult to find the expressions for $P_{L}(t)$ and $P_{R}(t)$ for this particular choice of $\mathbb{A}$. We omit these, as they are not particularly illuminating. Let us just point out that in all instances $P_{R}$ and $P_{L}$ are oscillating functions of time, although they are  generally not periodic. If one wishes to describe periodic Rabi oscillations in the $R$ and $L$ states, one should take, instead of (\ref{simple.1}),
\begin{equation}
\mathbb{A}=\frac{1}{2}\left(\begin{array}{cc}
1&-1\\
1&-1\\
\end{array}\right)
\end{equation}
as the natural choice for $\mathbb{A}$. By doing this one obtains expressions completely analogous to (\ref{Rah.1}) and (\ref{Rah.2}) for $P_{L}$ and $P_{R}$:
\begin{equation}
P_{L}(t)=\bigg(\frac{R_{1}}{R_{0}^{\prime}}\bigg)^{2}\sin^{2}\big(R_{0}^{\prime} t\big)
\end{equation}
\begin{equation}
P_{R}(t)=1-P_{L}(t),
\end{equation}
where the new frequency of the oscillation is now given by
\begin{equation}
R_{0}^{\prime}=\sqrt{\big(A/\hbar\big)^{2}+\big(\omega^{\prime}/2\big)^{2}}
\end{equation}
The main conclusion of this subsection is thus, that the Leggett-Caldeira can survive the influence of an environment on the particle in a SDWP under certain circumstances.  
\subsection{A limit on temperature}
An important result from perturbation theory tells us that for harmonic perturbations the time-dependent transition amplitude, $c_{n\rightarrow m}(t)$, between two given eigenstates of the complete Hamiltonian $H$ is given by:\cite{EspositoMarmoSudarshan}
\begin{equation}\label{III.6}
c_{n\rightarrow m}(t)=\langle \psi_{m}\vert A\vert\psi_{n}\rangle\frac{1-\exp i(\frac{E_{m}-E_{n}}{\hbar}-\omega^{\prime})t}{E_{m}-E_{n}-\hbar\omega^{\prime}}+\langle \psi_{m}\vert A^{*}\vert\psi_{n}\rangle\frac{1-\exp i(\frac{E_{m}-E_{n}}{\hbar}+\omega^{\prime})t}{E_{m}-E_{n}+\hbar\omega^{\prime}}.
\end{equation}
As a consequence we get that, if the system is to stay in the two lowest lying levels, then the perturbation must meet the condition:
\begin{equation}
\omega^{\prime}<\frac{E_{2}-E_{1}}{\hbar}.
\end{equation}
Otherwise, the perturbation would excite the system to higher levels with non-negligible probability.  This gives a limit on the temperature at which the system behaves like a low-lying two-level system. Indeed, recalling Wien's law for blackbody radiation, we get that thermal radiation at a temperature $T$ will have a maximal contribution of frequency $\omega^{\prime}$ when condition
\begin{equation}
\omega^{\prime}=\frac{2\pi c}{b_{W}}T
\end{equation}
is met. (Here, $T$ stands for the temperature of the radiation, $b_{W}$ is Wien's constant, and $c$ the velocity of light) Thus, if $\mathbb{V}(x,t)$ is somehow to represent thermal radiation, and if the perturbed Hamiltonian $H^{\prime}$ is to be described as a low-lying two-level system, then we must have:
\begin{equation}\label{temp.1}
T<\frac{b_{W}}{2\pi c}\ \frac{E_{2}-E_{1}}{\hbar}.
\end{equation}
In so many words: for each system there is a limit temperature above which the two-level system description is inapplicable, and zero-point  Leggett-Caldeira oscillations become overshadowed by other transitions. Moreover, from inequalities (\ref{inek01}) and (\ref{2NF}) we get :
\begin{equation}\label{temp.2}
T_{B}(a,m)<\frac{b_{W}}{2\pi c}\ \frac{E_{2}-E_{1}}{\hbar}<3T_{B}(a,m)
\end{equation}
with this global bound given by:
\begin{equation}\label{temp.3}
T_{B}(a,m)=\frac{5\pi\hbar b_{W}}{16mca^{2}}
\end{equation}
The meaning of expressions (\ref{temp.2}) and (\ref{temp.3}) is the following: consider a family of  double rectangular barriers, with a fixed $m$, $a$ and $k$, but free barrier width. When exposed to thermal radiation, there is a temperature $T_{B}$ for the radiation above which the Leggett-Caldeira oscillations are overshadowed by other transitions in least some the systems, and at temperature $3T_{B}$ the Calderia-Legget oscillations are surpassed by other transitions in all of the systems.
%
%
%%%%%%%%%%%%%%%%%%%%%%%%%%%%%%%%%%%%%%%%%%%%%%%%%%%%%%%%%%%%%%%%%%%%%%%%%
%
%
\section{Decoherence and dissipation}\label{sec:5}
We begin this section with a simplified exposition of mixed and pure states and the density matrix formalism as found in Landau and Lifshitz,\cite{LandauZ} to move on next to an also simplified rendering of some of Leggett's original argumentation. After that, decoherence is defined, an its relation with dissipation is briefly discussed.  

The interaction of a system ($\mathfrak{S}$) with its surroundings ($\mathfrak{E}$) can be taken into account by considering a bigger \emph{isolated} system ($\mathfrak{U}$) which encompasses both $\mathfrak{S}$ and $\mathfrak{E}$ (that is: $\mathfrak{U}=\mathfrak{S}\bigcup\mathfrak{E}$).  The state of this new, all including, system, $\mathfrak{U}$ is described by a state function $\Psi(j,\xi)$ that depends both on the coordinates of $\mathfrak{S}$  (the $j$) and the the coordinates of its environment (the $\xi$). The total Hamiltonian $H_{T}$ acting on $\mathfrak{U}$ can always be written in the form:
\begin{equation}
H_{T}=H+H_{\mathfrak{E}}+\lambda H_{I}
\end{equation}
where $H$ depends only on the $j$ and their generalized momenta, $H_{\mathfrak{E}}$ depends only on the $\xi$ and its momenta, and $H_{i}$ depends on both types of coordinates. We shall take the approximation, that $H$ is the Hamiltonian of $\mathfrak{S}$ when isolated, and that $H_{I}$ alone models the interaction between $\mathfrak{S}$ and $\mathfrak{E}$. 

In principle, there can happy instances in which $\Psi(j,\xi)$ could be written as the product of two states functions:
\begin{equation}\label{pure}
\Psi(j,\xi)=\psi(j)\phi(\xi)
\end{equation}
but this does not need to be the case. States that can be written in the form (\ref{pure}) are called \emph{pure states} in the literature. States that are not pure are said to be \emph{mixed.}

In order to illustrate this let us consider the case in which both the original system and its surroundings can be represented as two-level systems. If the isolated Hamiltonian $H$ has eigenfunctions $\psi_{+}$ and $\psi_{-}$:
\begin{equation}
H\psi_{\pm}=E_{\pm}\psi_{\pm}
\end{equation}
 and if $\phi_{\alpha}$ and $\phi_{\beta}$ are the eigenfunctions of $H_{e}$, \emph{i. e.} 
\begin{equation}
H_{e}\phi_{\alpha}=E_{\alpha}\phi_{\alpha}\ , \  H_{e}\phi_{\beta}=E_{\beta}\phi_{\beta},
\end{equation} 
then some examples of pure states are:
$$\begin{array}{ c }
\frac{1}{\sqrt{2}}(\psi_{+}\phi_{\beta}+\psi_{-}\phi_{\beta})=\frac{1}{\sqrt{2}}(\psi_{+}+\psi_{-})\phi_{\beta}, \ \frac{1}{2}\psi_{-}\phi_{\beta}+\frac{\sqrt{3}}{2}\psi_{-}\phi_{\alpha}=\psi_{-}(\frac{1}{2}\phi_{\beta}+\frac{\sqrt{3}}{2}\phi_{\alpha})\\
\textrm{and}\\ 
\frac{1}{4}\psi_{-}\phi_{\beta}+\frac{\sqrt{3}}{4}\psi_{-}\phi_{\alpha}-\frac{\sqrt{3}}{4}\psi_{-}\phi_{\beta}-\frac{3}{4}\psi_{+}\phi_{\alpha}=(\frac{1}{2}\psi_{-}-\frac{\sqrt{3}}{2}\psi_{+})(\frac{1}{2}\psi_{\beta}+\frac{\sqrt{3}}{2}\phi_{\alpha}).\\
\end{array}$$
On the other hand, as instances of mixed states, we can provide the following:
\begin{equation}\label{mixed}
\frac{1}{\sqrt{2}}(\psi_{-}\phi_{\alpha}+\psi_{+}\phi_{\beta})\  , \  \frac{1}{\sqrt{2}}(\psi_{-}\phi_{\beta}+\psi_{+}\phi_{\alpha})\ , \ \textrm{ and } \frac{1}{\sqrt{3}}(\psi_{+}\phi_{\alpha}+\psi_{+}\phi_{\beta}+\psi_{-}\psi_{\alpha}).
\end{equation}
The density matrix formalism was developed to treat systems that (like $\mathfrak{U}$) can present mixed states. The density matrix $\rho$ allows us to calculate the expected value $\langle f \rangle$ of any observable $f(x,p_{x})$ that depends only on the coordinates and momenta  of $\mathfrak{S}$:
\begin{equation}\label{Tr}
\langle f \rangle\ =\ \textrm{Tr}\Big( f\rho\Big). 
\end{equation} 
The elements of the density matrix $\rho$ of a state $\Psi(j,\xi)$ of are defined as:
\begin{equation}\label{rho}
\rho_{j,j^{\prime}}=S_{\xi}\Psi^{*}(j,\xi)\Psi(j^{\prime},\xi),
\end{equation} 
where $S_{\xi}$ stands for the sum over the discrete $\xi$ (if any) plus an integral over the continuous $\xi$ (if any). In the case of our $2\times 2$-level system, expression (\ref{rho}) reduces to 
\begin{equation}\label{rho}
\rho_{j,j^{\prime}}=\Psi_{j,\alpha}^{*}\Psi_{j^{\prime},\alpha}+\Psi_{j,\beta}^{*}\Psi_{j^{\prime},\beta}, \quad j,j^{\prime}=\pm\ .
\end{equation}
The diagonal elements of density matrix, of the form $\rho_{j,j}$, are called \emph{populations,} while the off-diagonal elements (\emph{i. e.} the elements with  $j\neq j^{\prime}$) are known as \emph{coherences.}

Suppose now that the 50-50 linear combinations 
\begin{equation}
\psi_{L}=\frac{1}{\sqrt{2}}\Big(\psi_{+}+\psi_{-}\Big), \ \psi_{R}=\frac{1}{\sqrt{2}}\Big(\psi_{+}-\psi_{-}\Big)
\end{equation}
are eigenfunctions of a macroscopic observable $M$, let us say:
\begin{equation}
M\psi_{L,R}=\mu_{L,R} \psi_{L,R},
\end{equation}
and take then the mixed state given by 
\begin{equation}\label{momix}
\Psi=c_{L}\psi_{L}\phi_{\alpha}+c_{R} \psi_{R}\phi_{\beta}.
\end{equation}
The density matrix associated with (\ref{momix}) is written as
\begin{equation}\label{diag}
\rho=\left(\begin{array}{cc}
\vert c_{L}\vert^{2}& 0\\
0&\vert c_{R}\vert^{2}\\
\end{array}\right)
\end{equation}
in the $\{\psi_{L}, \psi_{R}\}$ basis, as can be seen from (\ref{rho}), so that according to (\ref{Tr}) the expected value of any observable $f$ pertaining to $\mathfrak{S}$ yields the value
\begin{equation}\label{expvalue}
\langle f\rangle=\vert c_{L}\vert^{2}f_{L}+\vert c_{R}\vert^{2}f_{R},
\end{equation}
where $f_{L}$ and $f_{R}$ are the expected values of $f$ in the pure states
\begin{equation}
\Psi_{L}=\psi_{L}\phi_{\alpha}\ \textrm{ and }\ \Psi_{R}=\psi_{R}\phi_{\beta}. 
\end{equation}
The point of this discussion is that the same result (\ref{expvalue}) is obtained if we make measurements on an ensemble of $\mathfrak{U}$ systems all in state $\Psi$, or if the same measurements are made with an ensemble of $\mathfrak{U}$ made up of a statistical mixture of the pure states $\Psi_{L}$ and $\Psi_{R}$, in proportions $\vert c_{L}\vert^{2}$ and $\vert c_{R}\vert^{2}$. If it were to be held true that only ensembles of the type (\ref{momix}) could be prepared for $\mathfrak{U}$, then it could be argued that property $M$ has a sharp value for each element of the ensemble, and that a measurement done on a particular element only removes our ignorance on its value for that particular system. Clearly, this opens the door for hidden variable theories. To put it succinctly: in this interpretation each one of the Schroedinger's cats in an ensemble of such felines would be either dead or alive, and never in superpositions composed of both dead and alive states. Only the behaviour of the ensemble would be quantal, its individual elements being essentially classical.

It is patent, on the other hand, that the pure state
\begin{equation}
\Psi_{+}=\psi_{+}\phi_{\alpha}
\end{equation}  
cannot be written as a mixed state of the form (\ref{momix}) and that its corresponding density matrix cannot be diagonal in the $(L,R)$ basis, unlike (\ref{diag}). The impossibility of the simultaneous diagonalization of the density matrices of all possible states of a system $\mathfrak{S}$  is then a strong evidence of the true quantal behaviour of such system, as opposed to the behaviour required by hidden variable theories. Thus, for a system $S$ to be classical in any sense of the word, the coherences, \emph{i. e.} the off-diagonal elements, must be absent from the density matrix for each one of its possible states. This conclusion is generally valid, even if we resorted to the most trivial case in order to illustrate it.\cite{Leggett}

Decoherence can be defined as the decay of the off-diagonal elements in the density matrix as a result of the interaction of the system with its environment. Therefore decoherence allows a system to behave as quantal when isolated and as classical when the coupling with its environment is ``sufficiently effective." This is  nowadays considered a plausible mechanism for the emergence of classical reality from a quantal substratum.

In most practical applications, the environment $\mathfrak{E}$ has a very large number of degrees of freedom (say of the order of the Avogadro number) and not just one, as in the example we have used. Thus $\mathfrak{U}$ is usually a thermodynamic system, so that the full toolbox of quantum statistical mechanics needs to be marshalled in order to describe its behaviour. In this case the interaction between $\mathfrak{S}$ and  $\mathfrak{E}$ (interaction known as quantum dissipation in this context) involves the relaxation of the thermodynamical variables of $\mathfrak{U}$ towards thermal equilibrium, and not only decoherence.

Various models have been proposed over the years for the environment (or \emph{bath}) but one of first and most successful is the \emph{spin-boson Hamiltonian} approach, in which $H_{\mathfrak{E}}$ is taken as a collection of harmonic oscillators with various frequencies and the interaction term $H_{I}$ is linear both in the $j$ and in the $\xi$ coordinates. One important result from this approach is that a two-level system $\mathfrak{S}$ will describe damped oscillations between the localized states $\vert R\rangle$ and $\vert L\rangle$. Depending on the frequency distribution of the environment, $\mathfrak{S}$  may be localized at $T=0^{o}$K (the overdamped case, known as ``subohmic"), it may present critical damping (the ``ohmic case") or it may undergo underdamped coherent oscillations (the ``superohmic case.") The last one of these three instances is the most interesting for the present discussion, as it allows the observation of MQC before the complete relaxation of the system. The possibility of experimental MQC in the superohmic case depends in the interplay between a \emph{decoherence time} defined only by the bath parameters, and the period of the Leggett-Caldeira oscillation for system $\mathfrak{S}$.  
%%%%%%%%%%%%%%%%%%%%%%%%%%%%%%%%%%%%%%%%%%%%%%%%%%%%%%%%%%%%%%%%%%%%%%%%%%%
%
%
%                                                                                             Section 6 begins
%    
%
%%%%%%%%%%%%%%%%%%%%%%%%%%%%%%%%%%%%%%%%%%%%%%%%%%%%%%%%%%%%%%%%%%%%%%%%%%
\section{The scales of MQC}\label{sec:6}
\begin{table}[h!]
\centering
\begin{tabular}{|c|c|c|c|c|}
\hline
 $b$ & $E_0$ & $E_1$  & $\Delta E$ & $\tau$ \\
( nm )& $( \times 10^{-26} $ J )& ( $\times 10^{-26}$ J ) & $ ( \times 10^{-28}$J )&  (    $\mu$s   )\\
\hline \hline
100.00000& 5.3753895& 5.4382093& 6.3 &    1.0\\ \hline
116.65290& 5.3899569& 5.4246062& 3.5 &   2.9 \\  \hline
136.07900& 5.3987829& 5.4160961& 1.7 &   3.8\\ \hline
158.74011& 5.4036276& 5.4113353& 0.77 &   8.6\\ \hline
185.17494& 5.4059909& 5.4089897& 0.30 &   22.0\\ \hline
216.01195& 5.4069931& 5.4079902& 0.10 &   66.0\\ \hline
251.98421& 5.4073539& 5.4076298& 2.7$\times 10^{-2}$ & 240\\ \hline
\hline
\end{tabular}
\caption{Period $\tau$ increases exponentially as the barrier width is augmented. $a=1.0\ \mu$m, $k=2\times10^{-20}$J. This table, as well as all figures, was generated with Matlab \textsuperscript{\textregistered} R2012a.}
\label{tab:sample}
\end{table}
Let us start by fixing the width of the lateral wells at:
\begin{equation}\label{VII.1}
a=1\mu\textrm{m,}%a=1.0\times 10^{-8}\textrm{m}=10\textrm{nm},
\end{equation}
a value typical of contemporary lithographic circuitry, and take $m$ to be the rest mass of an electron:
\begin{equation}\label{VII.2}
m=m_{e}=9.1\times 10^{-31}\textrm{kg}.
\end{equation}
With this, $B$ takes the value:
\begin{equation}\label{5.A}
B=0.6\times 10^{-25}\textrm{J}=0.36\ \mu\textrm{eV,}
\end{equation}
and $T_{B}$ is fixed at:
\begin{equation}
T_{B}\approx 1.1\ \textrm{mK.}
\end{equation}

From equation (\ref{II.7}), that gives the fundamental frequency of the Caldeira-Leggett oscillations, we get the corresponding period 
\begin{equation}
\tau=\frac{2\pi\hbar}{E_{1}-E_{0}}.
\end{equation}
A global lower bound for this period is found from expressions (\ref{inequalityA}) and (\ref{funnyeq}):
\begin{equation}\label{5.B}
\tau >\frac{2\pi\hbar}{B}=\frac{4ma^{2}}{\pi\hbar}.
\end{equation}
For values (\ref{VII.1}) and (\ref{VII.2}) this gives
\begin{equation}
\tau>11\textrm{ns}.
\end{equation}
From table I (obtained through computer assisted numerical analysis) we get that as we sweep the barrier width from 0.2 to 0.5 $\mu$m the period of the Leggett -Caldeira oscillations for our square double well increases from 1.0 to 240 $\mu$s. Based on general considerations it has been estimated\cite{Leggett} that, for all practical purposes, MQC is lost if  the period of the Legget-Caldiera oscillation is of the order $\tau\gtrsim100\mu$s. Thus, the last row of the table corresponds to a localized system. All the other tabulated values could in principle correspond to observable MQC. 
\subsection{Some of the many things we have left out}
 MQC experiments are carried out in superconducting quantum interference devices (SQUIDs) with low capacitance tunneling Josephson junctions,\cite{Leggett, Friedmanetal} and the relevant coordinate (\emph{i. e.} the analogous of coordinate $x$) is not of a geometric character (like a position)  but is in most cases the phase difference between the states functions of the electrons in a Cooper pair (so that $m$ is not really the mass of the electron.) Thus our toy model is in reality a simplification of a mechanical analogy used to discuss experimental MQC. 
%
%
%
%%%%%%%%%%%%%%%%%%%%%%%%%%%%%%%%%%%%%%%%%%%%%%%%%%%%%%%%%%%%%%%%%%%%%%%%%%%%
 %
%
%
\section{Conclusions}\label{sec:7}
Contemporary quantum mechanics, both experimental and theoretical, provides examples of basic concepts and techniques such as: tunneling, stationary states, two-level systems, perturbation theory, the density matrix and the WKB approximation. Classroom presentations of current areas of research, such as MQC, help to improve the understanding of quantum physics at university level, as they connect the simplified textbook models with the actual state of the field, and thus with the future professional activity of the student. Moreover, MQC illustrates in a beautiful way the interplay between theory and experiment, and between concepts and techniques arising in different areas of quantum physics.   

We believe to have achieved in the present paper a level of exposition that makes it both clear and interesting for senior university students and recent graduates. In order to do so, we had to glide over the more technical aspects of experimental MQC and the intricate  relation between MQC and the epistemology and the philosophy of physics. We hope that the present paper will encourage the interested reader to delve further into this facets of contemporary research.
\section{Appendix}
Consider condition (\ref{coneven}) for the ground level ($n=0$), that is:
\begin{equation}\label{app01}
E_{0}\cot^{2} a\frac{\sqrt{2mE_{0}}}{\hbar}=(k-E_{0})\tanh^{2}b\frac{\sqrt{2m(k-E_{0})}}{\hbar}
\end{equation}
We will now establish a lower bound for $E_{0}$ starting from (\ref{app01}), but we have to take some precautions in doing so because $E_{0}$ depends implictly on $b.$ In order to proceed, note that
\begin{equation}\label{app01A}
\forall b\in(0,\infty)\ , \  \quad\frac{\sqrt{2m(k-E_{0})}}{\hbar}<\frac{\sqrt{2m(k-B/4)}}{\hbar}
\end{equation}
so that
\begin{equation}\label{app02}
\forall b\in(0,\infty) \ , \ \quad \tanh^{2}b\frac{\sqrt{2m(k-E_{0})}}{\hbar}>\tanh^{2} b\frac{\sqrt{2m(k-B/4)}}{\hbar}
\end{equation}
The dependence of the rhs of inequality (\ref{app02}) is explicit, so that the usual procedures of calculus can be applied. In particular as we now from elementary theorems that the limit 
\begin{equation}
\lim_{b\rightarrow \infty}\tanh^{2} b\frac{\sqrt{2m(k-B/4)}}{\hbar}=1
\end{equation}
holds true, we can affirm that: for given $\delta>0$ there exists a $b_{0}( \delta)$ such that any $b>b_{0}(\delta)$
\begin{equation}\label{app03}
\tanh^{2} b\frac{\sqrt{2m(k-B/4)}}{\hbar}>1-\frac{\delta}{2k}
\end{equation}
From (\ref{app01}) (\ref{app02}) and (\ref{app03}) we deduce that for any $b$ above a certain value $b_{0}(\delta)$, the ground energy of $U_{b}$ satisfies:
\begin{equation}\label{app04}
E_{0}\cot^{2}a\frac{\sqrt{2mE_{0}}}{\hbar}>(k-E_{0})\Big(1-\frac{\delta}{2k}\Big)
\end{equation}
Turning our attention to the condition for $E_{1}$, \emph{i. e.}
\begin{equation}\label{app01B}
E_{1}\cot^{2} a\frac{\sqrt{2mE_{1}}}{\hbar}=(k-E_{1})\coth^{2}b\frac{\sqrt{2m(k-E_{1})}}{\hbar}
\end{equation}
we now find an upper bound for $E_{1}$, by noting that, because of (\ref{app01A}) and the known properties of the hyperbolic functions, the inequality  
\begin{equation}
\coth^{2}b\frac{\sqrt{2m(k-E_{1})}}{\hbar}<\coth^{2}b\frac{\sqrt{2m(k-B/4)}}{\hbar}
\end{equation}
is verified for all strictly positive $b$. Furthermore,
\begin{equation}
\lim_{b\rightarrow\infty}\coth^{2}b\frac{\sqrt{2m(k-B/4)}}{\hbar}=1
\end{equation}
so that for every $\delta>0$ there exist a $b_{1}(\delta)$ such that, if  $b>b_{1}(\delta)$, then inequality
\begin{equation}\label{app07A}
\coth^{2}b\frac{\sqrt{2m(k-B/4)}}{\hbar}<1+\frac{\delta}{2k}
\end{equation}
is satisfied for all strictly positive $b.$ And from (\ref{app01B}) and (\ref{app07A}) we get that, for all $b$ above a certain thershold value $b_{1}(\delta)$, the inequality 
\begin{equation}\label{app08}
E_{1}\cot^{2}a\frac{\sqrt{2mE_{1}}}{\hbar}<(k-E_{1})\Big(1+\frac{\delta}{2k}\Big)
\end{equation}
is satisfied.

Taking both (\ref{app04}) and (\ref{app08}) into consideration, we have that for every $\delta >0$ there exists a number $b(\delta)=\max \{b_{0}(\delta) b_{1}(\delta)\}$ such that for any $b>b(\delta)$ the inequality
\begin{equation}\label{app09}
E_{1}\cot^{2}a\frac{\sqrt{2mE_{1}}}{\hbar}-E_{0}\cot^{2}a\frac{\sqrt{2mE_{0}}}{\hbar}<(E_{0}-E_{1})+\delta \Big(1-\frac{E_{0}+E_{1}}{2k}\Big)
\end{equation}
is satisfied. Now, it is not difficult to see that
\begin{equation}
w(E)=E\cot^{2}a\frac{\sqrt{2mE}}{\hbar}
\end{equation} 
is a monotonically increasing function of $E$ in the range $B/4<E<B$, so that 
\begin{equation}\label{app010}
0<E_{1}\cot^{2}a\frac{\sqrt{2mE_{1}}}{\hbar}-E_{0}\cot^{2}a\frac{\sqrt{2mE_{0}}}{\hbar}
\end{equation} 
and in the other hand, we deduce
\begin{equation}\label{app011}
(E_{0}-E_{1})+\delta \Big(1-\frac{E_{0}+E_{1}}{2k}\Big)<\Big(1-\frac{E_{0}+E_{1}}{2k}\Big)\delta <\delta 
\end{equation}
from and . From ( \ref{app09}), (\ref{app010}) and (\ref{app011}) we get:
\begin{equation}
0<E_{1}\cot^{2}a\frac{\sqrt{2mE_{1}}}{\hbar}-E_{0}\cot^{2}a\frac{\sqrt{2mE_{0}}}{\hbar}<\delta
\end{equation}
\begin{equation}
E_{1}\cot^{2} a\frac{\sqrt{2mE_{1}}}{\hbar}>k-E_{1}
\end{equation}
Finally, we notice that, as 
\begin{equation}
v(E)=\cot^{2}a\frac{\sqrt{2mE}}{\hbar}
\end{equation}
is a monotonically increasing function of $E$ in the range $B/4<E<B$, then 
\begin{equation}\label{app013}
(E_{1}-E_{0})\cot^{2}a\frac{\sqrt{2mE_{0}}}{\hbar}<\delta
\end{equation}
Now we just need to find a lower bound on $\cot^{2}a\frac{\sqrt{2mE_{0}}}{\hbar}$. This is obtained by turning back to condition (\ref{app01}) from which we get
\begin{equation}\label{app014}
\cot^{2}a\frac{\sqrt{2mE_{0}}}{\hbar}<\frac{k-B/4}{B}
\end{equation}
Finally, from (\ref{app013}) and (\ref{app014}) we arrive at
\begin{equation}\label{app015}
E_{1}-E_{0}<\delta\frac{B}{k-B/4}
\end{equation}
Let us stress that $k$ and $B$ are independent of $b$. In this manner, we have arrived at the following lemma:

For each strictly positive real number $ \delta$ there exists a
\begin{equation}
b^{\prime}(\delta)=b(\delta\frac{k-B/4}{B})
\end{equation}
such that for any $b>b^{\prime}(\delta)$ the gap between the ground and first excited levels of of $U_{b}$ is less than $\delta$, that is, such that:
\begin{equation}
E_{1}-E_{0}<\delta \ .
\end{equation}
And this is what we set out to prove in this appendix.

\end{document}